\documentclass{emulateapj}

\begin{document}

\title{Axions and the cooling of white dwarf stars}

\author{J. Isern$^{1,2}$, 
        E. Garc\'{\i}a--Berro$^{3,2}$, 
        S. Torres$^{3,2}$, S. Catal\'an$^{1,2}$} 

\altaffiltext{1}{Institut de Ci\`encies de l'Espai, CSIC, 
                 Facultat de Ci\`encies, Campus UAB, 
                 08193 Bellaterra, Spain}
\altaffiltext{2}{Institut d'Estudis Espacials de Catalunya, 
                 c/ Gran Capit\`{a} 2--4, 08034 Barcelona, Spain}
\altaffiltext{3}{Departament de F\'isica Aplicada, 
                 Escola Polit\`ecnica Superior de Castelldefels, 
                 Universitat Polit\`ecnica de Catalunya,
                 Avda. del Canal Ol\'\i mpic s/n, 
                 08860 Castelldefels, Spain}

\email{isern@ieec.fcr.es}

\begin{abstract}
White dwarfs are the end product of the lifes of intermediate- 
and low-mass stars and their evolution is described as a simple 
cooling process.  Recently, it has been possible to determine 
with an unprecedented precision their luminosity function, that 
is, the number of stars per unit volume and luminosity interval. 
We show here that the shape of the bright branch of this function 
is only sensitive to the averaged cooling rate of white dwarfs 
and we propose to use this property to check the possible 
existence of axions, a proposed but not yet detected weakly 
interacting particle.  Our results indicate that the inclusion of 
the emission of axions in the evolutionary models of white dwarfs 
noticeably improves the agreement between the theoretical 
calculations and the observational white dwarf luminosity 
function. The best fit is obtained for $m_a \cos^2 \beta 
\approx 5$~meV, where $m_a$ is the mass of the axion and 
$\cos^2 \beta$ is a free parameter. We also show that values 
larger than 10 meV are clearly excluded.  The existing 
theoretical and observational uncertainties do not yet allow the confirmation 
of the existence of axions, but our results clearly show 
that if their mass is of the order of few meV, the white dwarf 
luminosity function is sensitive enough to detect their 
existence. \end{abstract}

\keywords{elementary  particles  --- stars:  luminosity function, mass function ---  white dwarfs}

\section{Introduction}

One solution to the strong CP problem of quantum chromodynamics 
is the Peccei-Quinn symmetry (Peccei \& Quinn 1977a, 1977b).  
This symmetry is spontaneously broken at an energy scale that 
gives rise to the formation of a light pseudo-scalar particle 
named the ``axion'' \citep{wei78,wil78}.  This scale of energies 
is not defined by the theory but it has to be well above the 
electroweak scale to ensure that the coupling between axions and 
matter is weak enough to account for the lack of a positive 
detection up to now. The mass of axions and the energy scale are 
related by $m_a\approx0.6(10^7~{\rm GeV}/f_a)$~eV.  
Astrophysical and cosmological arguments \citep{raf07} have been 
used to constrain this mass to the range $10^{-2}~{\rm eV}\geq 
m_a \geq 10^{-4}~{\rm eV}$.  For this mass range, axions 
can escape from stars and act as a sink of energy.

White dwarfs are the final evolutionary phase of low- and 
intermediate-mass stars ($M\leq10\pm2~M_{\sun}$).  Since they are 
degenerate objects, they cannot obtain energy from thermonuclear 
reactions and their evolution can be described just as a 
gravothermal process of cooling.  Therefore, if axions exist, the 
properties of these stars would be noticeably perturbed.  
Furthermore, white dwarfs have a relatively simple structure:  a 
degenerate core that contains the bulk of the mass and acts as an 
energy reservoir and a partially degenerate envelope that 
controls the energy outflow.  The vast majority of white dwarfs 
have masses in the range $0.4\leq M/M_{\odot}\leq 1.05$ --- 
although these figures are still uncertain --- and have a core 
made of a mixture of carbon and oxygen.  All of them are 
surrounded by a thin helium layer, with a mass ranging from 
$10^{-2}$ to $10^{-4}~M_{\odot}$ which, in turn, is surrounded by 
an even thinner layer of hydrogen with a mass between $10^{-4}$ 
and $10^{-15}~M_{\sun}$, although about 25\% of white dwarfs do 
not have hydrogen atmospheres.  White dwarfs displaying hydrogen 
in their spectra are called DA and the remaining ones are known 
as non-DAs. Because of the different opacities, DA white dwarfs 
cool more slowly than the non-DA ones.

The standard theory of white dwarf cooling can be summarized as 
follows \citep{ise98}. When the luminosity is large, $M_{\rm 
bol}<8$, the evolution is dominated by neutrino emission.  In 
this phase the main uncertainties come from our poor knowledge of 
the initial conditions.  Fortunately, it has been shown that all 
the initial thermal structures converge toward a unique one 
\citep{dan89}. For smaller luminosities, $8\leq M_{\rm 
bol}\leq12$, the main source of energy is of gravothermal origin.  
In this phase, the Coulomb plasma coupling parameter is not large 
and the cooling can be accurately described.  Furthermore, the 
energy flux through the envelope is controlled by a thick 
nondegenerate or partially degenerate layer with an opacity 
dominated by hydrogen, when present, and helium, and it is weakly 
dependent on the metal content since metals sink towards the base 
of the envelope by gravitationally induced diffusion.  Below 
these luminosities, white dwarfs evolve into a region of 
densities and temperatures where the plasma crystallizes.  When 
this happens, two additional sources of energy appear. The first 
one is the release of latent heat during crystallization. The 
second one is the release of gravitational energy induced by 
phase separation of the different chemical species 
(Garc\'{\i}a-Berro et al.~1988a, 1988b; Isern et al.~1997, 2000).  When the bulk of the star is 
solid the white dwarf enters into the Debye cooling phase and the 
only important source of energy comes from the compression of the 
outer layers.  These late phases of cooling are not yet well 
understood \citep{ise98}.

\section{The white dwarf luminosity function}

\begin{figure}
\vspace{8cm}
\includegraphics{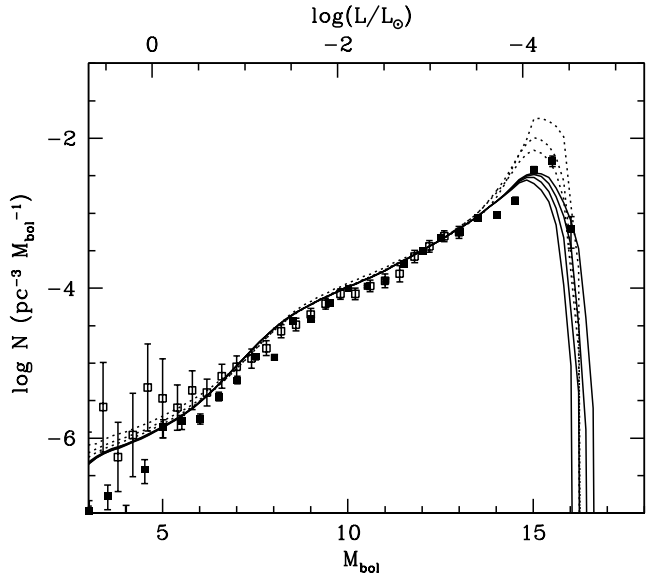}
\caption{Luminosity  functions   of  white  dwarfs.    Filled  squares
         correspond to the luminosity function of white dwarfs (DA and
         non-DA types) obtained using the reduced proper motion method
         \citep{har06}  and  open   squares  to  that  obtained  using
         spectroscopically  identified DA white  dwarfs \citep{deg08}.
         The  solid  lines are  the  theoretical luminosity  functions
         obtained for  different ages of  the Galaxy --- from  left to
         right:  10,  11,  12 and  13  Gyr  ---  and a  constant  star
         formation  rate.   The  dotted  lines  are the  same  for  an
         exponentially    decreasing   star    formation    rate   ---
         $\psi(t)=\exp{(-t/\tau)}$  with $\tau=0.5$, 3  and 5  Gyr ---
         and the same  age of the Galaxy, 11  Gyr.  All the luminosity
         functions have been normalized to the same observational data
         point.}
\label{axion1}
\end{figure}

One way to test the evolutionary properties of white dwarfs is 
using their luminosity function, which is defined as the number 
of white dwarfs per unit volume and magnitude. The first 
luminosity function was derived four decades ago \citep{wei68} 
and since then it has been noticeably improved. The most recent 
determinations use data from the Sloan Digital Sky Survey (SDSS).  
The first of these \citep{har06} was built from a sample of 6000 
DA and non-DA white dwarfs with accurate photometry and proper 
motions culled from the SDSS Data Release 3 and the USNO-B 
catalogue, whereas the second one \citep{deg08} was constructed 
from a sample of 3528 spectroscopically identified DA white 
dwarfs from the SDSS Data Release 4 (see Fig.~\ref{axion1}). 
The monotonic behavior of this function clearly proves that the 
evolution of white dwarfs is just a cooling process. The sharp 
cutoff at low luminosities is a consequence of the finite age of 
the Galactic disk.  The luminosity function of white dwarfs can 
be computed as:

\begin{equation}
n(M_{\rm bol})=\int_{M_{\rm l}}^{M_{\rm u}}
\phi(M) \psi(T_{\rm G}-t_{\rm cool}-t_{\rm ps})\tau_{\rm cool}\;dM,
\label{eq:wdlf}
\end{equation}

\noindent where $M_{\rm bol}$ is the bolometric magnitude, $M$ is 
the mass of the parent star --- for convenience all white dwarfs 
are labeled with the mass of its progenitor --- $T_{\rm G}$ is 
the age of the population under study, $t_{\rm cool}$ is the time 
that a white dwarf with a progenitor of mass $M$ takes to cool 
down to a bolometric magnitude $M_{\rm bol}$, $\tau_{\rm cool}$ 
is the characteristic cooling time of the white dwarf, $\tau_{\rm 
cool}= dt_{\rm cool}/dM_{\rm bol}$, and $t_{\rm ps}$ is the 
lifetime of the parent star.  In equation (\ref{eq:wdlf}) $M_{\rm u}$ 
is the maximum mass of white dwarf progenitors and $M_{\rm l}$ is 
the mass of the progenitor that satisfies the condition $T_{\rm 
G}-t_{\rm cool}-t_{\rm ps}=0$, i.e., the minimum mass of a star 
able to produce a white dwarf of the required luminosity.  The 
remaining quantities are the initial mass function, $\phi$ (here 
we have used Salpeter's law),  and the star formation rate, 
$\psi$, which is not known a priori and depends on the 
population under study.

Since neither the star formation rate nor the total density of 
white dwarfs are known, theoretical luminosity functions are 
normalized to a given observational bin, usually the one with the 
smallest error bar.  Figure \ref{axion1} displays the observed 
luminosity functions and several luminosity functions obtained 
with the same DA cooling sequences \citep{sal00} and different 
star formation rates and ages of the Galaxy.  The cooling models 
assume a nonhomogeneous distribution of carbon and oxygen in the 
core \citep{sal97}, a pure helium layer of $10^{-2} \, M_{*}$ and 
on top of it a pure hydrogen layer of $10^{-4}\, M_*$, where 
$M_*$ is the mass of the white dwarf. Since a relationship 
connecting the mass of the white dwarf and the mass of its 
progenitor is also necessary we have adopted the one that best 
reproduces the mass distribution of white dwarfs \citep{cat08}.

An interesting feature of Figure \ref{axion1} is that the bright 
part of the white dwarf luminosity function --- that with 
bolometric magnitude $M_{\rm bol}<13$ --- is almost independent 
of the assumed star formation rate.  This can be explained with 
simple arguments.  Since the characteristic cooling time is not 
strongly dependent on the mass of the white dwarf, 
equation (\ref{eq:wdlf}) can be written as

\begin{equation}
n\propto \left\langle\tau_{\rm cool}\right\rangle 
\int_{M_{\rm l}}^{M_{\rm u}}\phi(M)
\psi(T-t_{\rm cool}-t_{\rm ps})\;dM. 
\end{equation}

Restricting ourselves to bright white dwarfs --- namely, those 
for which $t_{\rm cool}$ is small --- the lower limit of the 
integral is satisfied by low-mass stars and, as a consequence of 
the strong dependence of the main-sequence lifetimes with mass, 
it takes a value that is almost independent of the luminosity 
under consideration. Therefore, if $\psi$ is a well-behaved 
function and $T_{\rm G}$ is large enough, the integral is not 
sensitive to the luminosity, its value is absorbed by the 
normalization procedure, and the shape of the luminosity function 
only depends on the averaged physical properties of white dwarfs.  
It is important to mention here that the initial-final mass 
relationship enters as a weight into the calculation of this 
average.  Neverheless, since only those functions able to provide 
a good fit to the mass distribution of white dwarfs are 
acceptable, its influence on the shape of the bright branch of 
the luminosity function is minor. Here we use this property of 
the white dwarf luminosity function, together with the recently 
obtained high-precision observational luminosity function, to 
study the influence of the emission of axions, and to check which 
mass of the axion is compatible with observations. The idea of 
using the cooling times of white dwarfs to constrain the 
properties of axions is not new \citep{raf86}, but the crudeness 
of the theoretical models and of the observational data prevented 
a definite conclusion and just a loose upper bound was obtained, 
$m_a<30$ meV.

\begin{figure}
\vspace{8cm}
\includegraphics{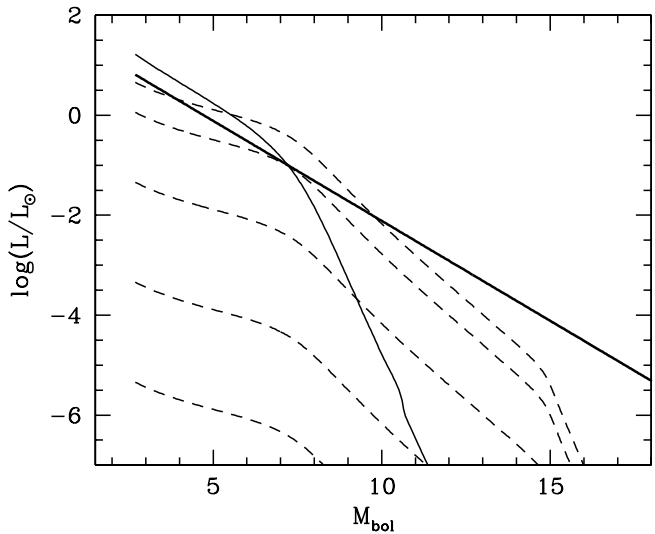}
\caption{Energy  losses for a  $ 0.61  \, M_{\sun}$  white dwarf  as a
         function  of  the  bolometric  magnitude.  The  dashed  lines
         represent  the  axion  luminosity  for  different  values  of
         $m_a \cos^2  \beta$ --- from top to  bottom: $m_a
         \cos^2 \beta$ =10,  5, 1, 0.1 and 0.01  meV.  The thick solid
         line  is the  photon luminosity,  while the  thin  solid line
         shows the neutrino luminosity.}
\label{axion2}
\end{figure}

\section{Axions and the white dwarf luminosity function}

Axions can couple to photons, electrons and nucleons with a 
strength that depends on the specific implementation of the 
Peccei-Quinn mechanism.  The two most common implementations are 
the KVSZ \citep{kim79,shi80} and the DFSZ models 
\citep{din81,zhi80}.  In the first case axions couple to hadrons 
and photons, while in the second they also couple to charged 
leptons.  For the temperatures and densities of the white dwarfs 
under consideration, only DFSZ axions are relevant and in this 
case they can be emitted by Compton, pair annihilation and 
bremsstrahlung processes, but only the last mechanism turns out 
to be important. Figure \ref{axion2} shows the energy losses for 
an otherwise typical $0.61 \, M_{\sun}$ white dwarf as a function 
of the bolometric magnitude. The dashed lines represent the axion 
luminosity for different values of $m_a \cos^2 \beta$. The 
axion emission rate (in erg~g$^{-1}$~s$^{-1}$)  has been computed 
\citep{nak87,nak88} as $\varepsilon_a=1.08\times10^{23} 
\alpha (Z^2/A)T_7^4 F$, where $F$ is a function of the temperature 
and the density which takes into account the properties of the 
plasma, and $\alpha={g_{ae}^2}/{4\pi}$ is related to the 
axion-electron coupling constant $g_{ae}=2.8\times10^{-11} 
m_a \cos^2 \beta /1~{\rm eV}$.  Since the core is almost 
isothermal it turns out that $L_a \propto T^4$ in the 
region in which axions are the dominant sink of energy. The thick 
solid line represents the photon luminosity \citep{sal00}. For 
the region of interest $L_\nu \propto T^a$, with $a\sim2.6$, 
although this value changes as the white dwarf cools down. The 
thin solid line represents the neutrino luminosity, which scales 
as $L_\nu \propto T^8$ and is also dominated by the plasma and 
bremsstrahlung processes \citep{ito96}.  Therefore, since the 
temperature dependence of the different energy-loss processes is 
not the same, the luminosity function allows to disentangle the 
different contributions.

\begin{figure}
\vspace{8cm}
\includegraphics{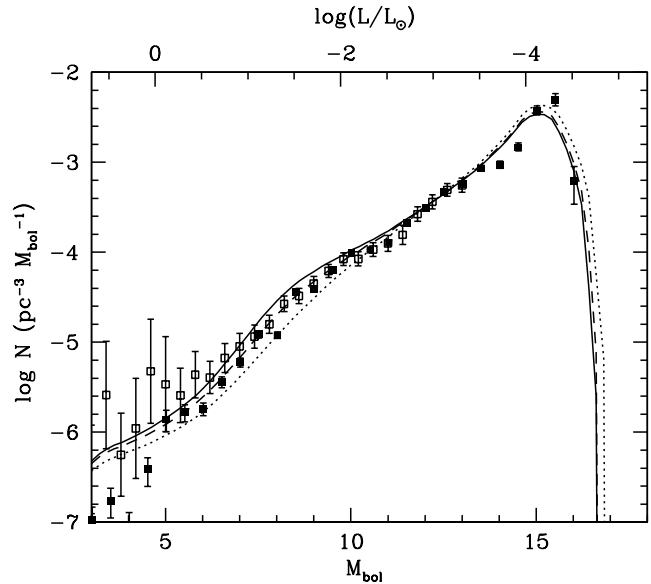}
\caption{White dwarf luminosity functions  for different values of the
         axion  mass.  The  luminosity  functions have  been  computed
         assuming $m_a \cos^2 \beta$ = 0 (solid line), 5 (dashed
         line) and 10 (dotted line) meV.}
\label{axion3}
\end{figure}

Figure \ref{axion2} shows that in the region $M_{\rm bol}\sim10$ 
the axion luminosity can be comparable with the photon and 
neutrino ones, depending on the adopted axion mass. It also shows 
that the region around $M_{\rm bol}\sim12$ provides a solid 
anchor point to normalize the luminosity function because there 
the observational data have reasonably small error bars, models 
are reliable, neutrinos are not relevant and axions, if they 
exist, are not dominant.

\section{Results and Conclusions}

\begin{figure}
\vspace{8cm}
\includegraphics{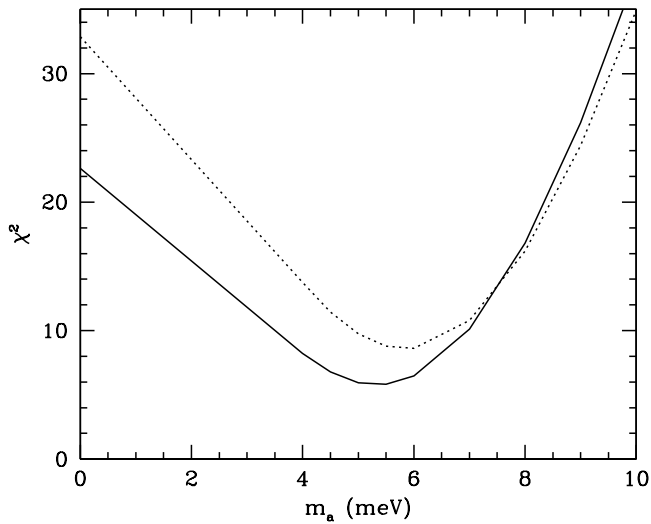}
\caption{Value of $\chi^2$ as a function  of the mass of the axion for
         the  case in  which the  initial--final mass  relationship of
         \cite{cat08}  (solid line)  and that  of  \cite{wood} (dotted
         line) are used.}
\label{axion4}
\end{figure}

Figure \ref{axion3} displays several luminosity functions 
obtained using different axion masses, adopting a constant star 
formation rate and an age of the Galactic disk of 11 Gyr.  As 
already mentioned, it is important to realize that the bright 
branch of the luminosity function is not sensitive to these last 
assumptions.  All the luminosity functions have been normalized 
to the luminosity bin at $\log (L/L_{\sun})\simeq-3$ or, 
equivalently, $M_{\rm bol} \simeq 12.2$.  The best-fit model --- 
namely that which minimizes the $\chi^2$ test in the region 
$-1>\log (L/L_{\sun})>-3$ (that is, $ 7.2 < M_{\rm bol} < 12.2$), 
which is the region where both the observational data and the 
theoretical models are reliable --- is obtained for $m_a 
\cos^2 \beta\approx 5$ meV and solutions with $m_a \cos^2 
\beta > 10$ meV are clearly excluded.  Figure \ref{axion4} 
displays the behavior of $\chi^2$ as a function of the mass of 
the axion in our fiducial case (solid line) and in the case in 
which we use the initial-final mass relationship of \cite{wood}, 
which is marginally compatible with the white dwarf mass 
distribution \citep{cat08}.  In both cases the behavior of the 
luminosity function is similar and gives similar values for the 
mass of the axion once one takes into account the present 
uncertainties.  It is also important to notice that the largest 
contribution to the lack of accuracy comes from the brightest 
bins of the luminosity function, which have large error bars.

This result is completely compatible with the previously existing 
constraints \citep{raf07}.  Furthermore, these values are also 
compatible with the bounds imposed by the drift of the 
pulsational period of the ZZ Ceti star G117$-$B15A 
\citep{ise92,cor01,bis08}.  It is worthwhile to mention here that 
axions with $m_a \approx 5$ meV would change the expected 
period drift of variable DB white dwarfs --- which have values 
between $\dot P \sim 10^{-13}$ and $10^{-14}$ s~s$^{-1}$ 
\citep{cor04} --- by a factor 2, the exact value depending on the 
adopted temperature of the stellar core and that the detection of 
such a drift would provide a strong additional argument in favor 
of the existence of axions.

The results presented here are not a definite proof of the 
existence of axions, since there are still some observational and 
theoretical uncertainties. However, the calculations reported 
here show that the hot branch of the white dwarf luminosity 
function is a powerful tool to test the existence of weakly 
interacting massive particles because it is only sensitive to the 
averaged cooling rate of white dwarfs and not to the details of 
the star formation rate or the initial mass function, as shown in 
\S 2.  Moreover, our results are problably the first evidence 
that the shape of the white dwarf luminosity function could be 
affected by the emission of axions and that this change can be 
measured. If this is indeed the case the mass of the axion would 
be of about 5 meV. In addition, we have derived an upper bound to 
the mass of the axions of 10 meV, which is compatible with other 
recent determinations \citep{bis08}.  It is worth 
mentioning that this result is relevant for other research fields 
and, in particular, for cosmology.  Specifically, assuming 
$\cos^2\beta = 1$, the contribution of axions to dark matter 
would be of the order of 0.2\% \citep{raf07}.  In view of 
this, we consider it of the largest importance to improve the 
observational determination of the white dwarf luminosity 
function, especially in the region of the hottest white dwarfs 
(see Fig.~\ref{axion3}).  Thus, the extension of the SDSS is of 
the maximum interest not only for astronomers and cosmologists, 
but also for particle physicists.  However, not only 
observational efforts are needed, since in order to obtain a 
reliable determination of the mass of the axion it is also 
important to decrease the uncertainties in the plasmon neutrino 
emission rates at the relevant temperature range.  Furthermore, 
it would be also convenient to intensify the study of the drift 
of the pulsational periods of variable white dwarfs in order to 
obtain additional independent evidence.

\acknowledgments Part of this work was supported by the MEC 
grants AYA05-08013-C03-01 and 02, and ESP2007-61593 by the 
European Union FEDER funds and by the AGAUR.  We thank M. Salaris 
for providing us with the neutrino emission files corresponding 
to the models of \citet{sal00}.

\end{document}